\documentclass[10.5pt]{IEEEtran}
\usepackage{graphicx}
\usepackage{amsmath}
\usepackage{multicol}
\usepackage{multirow}
\usepackage{stfloats}
\usepackage{booktabs}
\usepackage{mathrsfs}
\usepackage{enumerate}
\usepackage{amssymb}
\usepackage{algorithm}
\usepackage{algpseudocode}

\usepackage{array}
\usepackage{underscore}
\usepackage{breqn}
\usepackage{url}

\usepackage{flushend}

\usepackage[square, comma, sort&compress, numbers]{natbib}
\usepackage{graphicx}
\usepackage{epstopdf}
\usepackage{caption}
\usepackage{diagbox}

\usepackage{caption}
\usepackage{subfigure}
\usepackage{color}

\usepackage[outerbars,color]{changebar}
\ifx\pdfoutput\undefined
\else\ifnum\pdfoutput>0
  \usepackage{pdfcolmk}
\fi\fi
\cbcolor{red}
\usepackage{fancyhdr}

\makeatletter
\newcommand{\Rmnum}[1]{\expandafter\@slowromancap\romannumeral #1@}
\makeatother

\usepackage{array}  \newcommand{\PreserveBackslash}[1]{\let\temp=\\#1\let\\=\temp}  \newcolumntype{C}[1]{>{\PreserveBackslash\centering}p{#1}}  \newcolumntype{R}[1]{>{\PreserveBackslash\raggedleft}p{#1}}  \newcolumntype{L}[1]{>{\PreserveBackslash\raggedright}p{#1}}

\ifCLASSINFOpdf
\else
\fi
\begin{document}

\title{\begin{Huge}Channel-Dependent Scheduling in Wireless Energy Transfer for Mobile Devices\end{Huge}}

\author{Wen Fang,
Gang Wang,~\IEEEmembership{Member,~IEEE},
Georgios B. Giannakis,~\IEEEmembership{Fellow,~IEEE,}\\
Qingwen Liu,~\IEEEmembership{Senior Member,~IEEE},
Xin Wang,~\IEEEmembership{Senior Member,~IEEE},
and Hao Deng

	\thanks{W.~Fang, Q. Liu, and H. Deng are with the College of Electronic and Information Engineering, Tongji University, Shanghai 200000, China (e-mail: wen.fang@tongji.edu.cn, qliu@tongji.edu.cn, and denghao1984@tongji.edu.cn).

G. Wang and G. B. Giannakis are with the Digital Technology Center and the Department of Electrical and Computer Engineering, University of Minnesota, Minneapolis, MN 55455, USA (e-mail: gangwang@umn.edu, and georgios@umn.edu).

X.~Wang is with the School of Information Science and Technology, Fudan University, Shanghai 200000, China (e-mail: xwang11@fudan.edu.cn).

}
}

\maketitle

\begin{abstract}

\normalsize
Resonant Beam Charging (RBC) is a long-range, high-power, mobile, and safe Wireless Energy Transfer (WET) technology, which can provide wireless power for mobile devices like Wi-Fi communications. Due to the wireless energy transmission decay in RBC systems, the charging power received per device relies on the distance-dependent energy transmission channel. To extend battery life of all devices, this paper develops a Channel-Dependent Charge (CDC) scheduling algorithm to control receivers' charging power, order and duration. Each receiver is assigned a dynamic scheduling coefficient, which is the product of the battery's remaining energy and the energy transmission channel. The resultant optimal charging order is to charge the receiver with the minimum scheduling coefficient first per equal unit-length time slot. It is shown analytically and experimentally that the CDC algorithm achieves higher charging performance than other scheduling algorithms including the Round-Robin Charge (RRC) scheduling algorithms.
In a word, the CDC scheduling algorithm offers a viable approach to extending mobile devices' battery life while accounting for varying RBC transmission channels.

\end{abstract}


\IEEEpeerreviewmaketitle

\section{Introduction}\label{Section1}
Recent advances in wireless communications enable mobile devices to have become the necessity of our daily life \cite{sample2011analysis}. On the other hand, the Internet of Things (IoT), which holds the key to realizing the vision of a global infrastructure of networked physical objects and has attracted great attention recently, requires connecting multiple mobile devices \cite{zhao2015wireless, xia2012internet, chen2019Learning}. These mobile devices can be sensors, actuators, smartphones, computers, and other appliances that can be connected, monitored or actuated \cite{wu2014cognitive}. In particular, mobile devices are often powered by batteries to enable high-performance computing and communications \cite{wang2018power, guo2016minimizing, du2017contract}. However, current short battery endurance limits the widespread use of mobile devices \cite{guo2018energy}. The contradictions between the evolution of IoT and the battery endurance of mobile devices are becoming more and more prominent \cite{wang2018power, cheng20175g}. Current approaches to improving battery endurance include: i) increasing the battery capacity, and, ii) adopting new charging methods \cite{QingqingMET}.

Limited by chemical characteristics of ordinary batteries, increasing their capacity is faced with cost, technology, and other challenges \cite{panigrahi2001battery}. Carrying a charging cable or looking for a charging socket, however, brings inconvenience to mobile device users. Therefore, the natural energy harvesting methods and Wireless Energy Transfer (WET) technology have become two main strategies for overcoming the energy supply limitations \cite{chi2019energy}. The natural energy harvesting is the technology that harvests energy from natural sources of clean energy such as solar and wind \cite{sun2017coordinated}. However, the energy harvested in this way is unstable and easily affected by geography, environment and other factors \cite{zhao2017joint, deng2018multisource, li2017multiuser}. WET technology has developed rapidly in the last decade \cite{Hui2014, lu2016wireless}, and it is identified as the key technology solution for $6$G \cite{david20186g}. As a matter of fact, most existing WET technologies such as inductive coupling, magnetic resonance coupling, and radio frequency, face major issues related to e.g., charging safety, power, and distance \cite{wirelesstechniques, electromagnetic, cheng2016consumer}.

To provide safe, high-power, long-distance and mobile charging services for mobile devices, the Resonant Beam Charging (RBC), also known as Distributed Laser Charging (DLC), was recently advocated in \cite{liu2016dlc, fang2018}. An RBC system comprises two main components, named the transmitter and the receiver, and they are separated in space \cite{fang2018}. As long as the receiver is within Line of Sight (LOS) of the transmitter, a resonant beam (i.e., the transmission channel) can be generated between them to transfer power without alignment. With regards to safety, the resonant beam can be automatically cut off whenever an object enters the LOS blocking the transmission channel. With the high power and high collimation of the laser, the high-power and long-distance charging of the RBC system can be realized. Furthermore, multiple beams can be generated between an RBC transmitter and multiple receivers simultaneously. In other words, multiple mobile devices can be served by a transmitter simultaneously in an RBC system. As such, the RBC system is well-suited for providing devices with mobile, safe, and `WiFi-like' wireless charging services; see also \cite{liu2016dlc, Qing2017, fang2018, zhang2018distributed2, xiong2018tdma} for related discussion.
\begin{figure}[!t]
    \centering
    \includegraphics[scale=0.233]{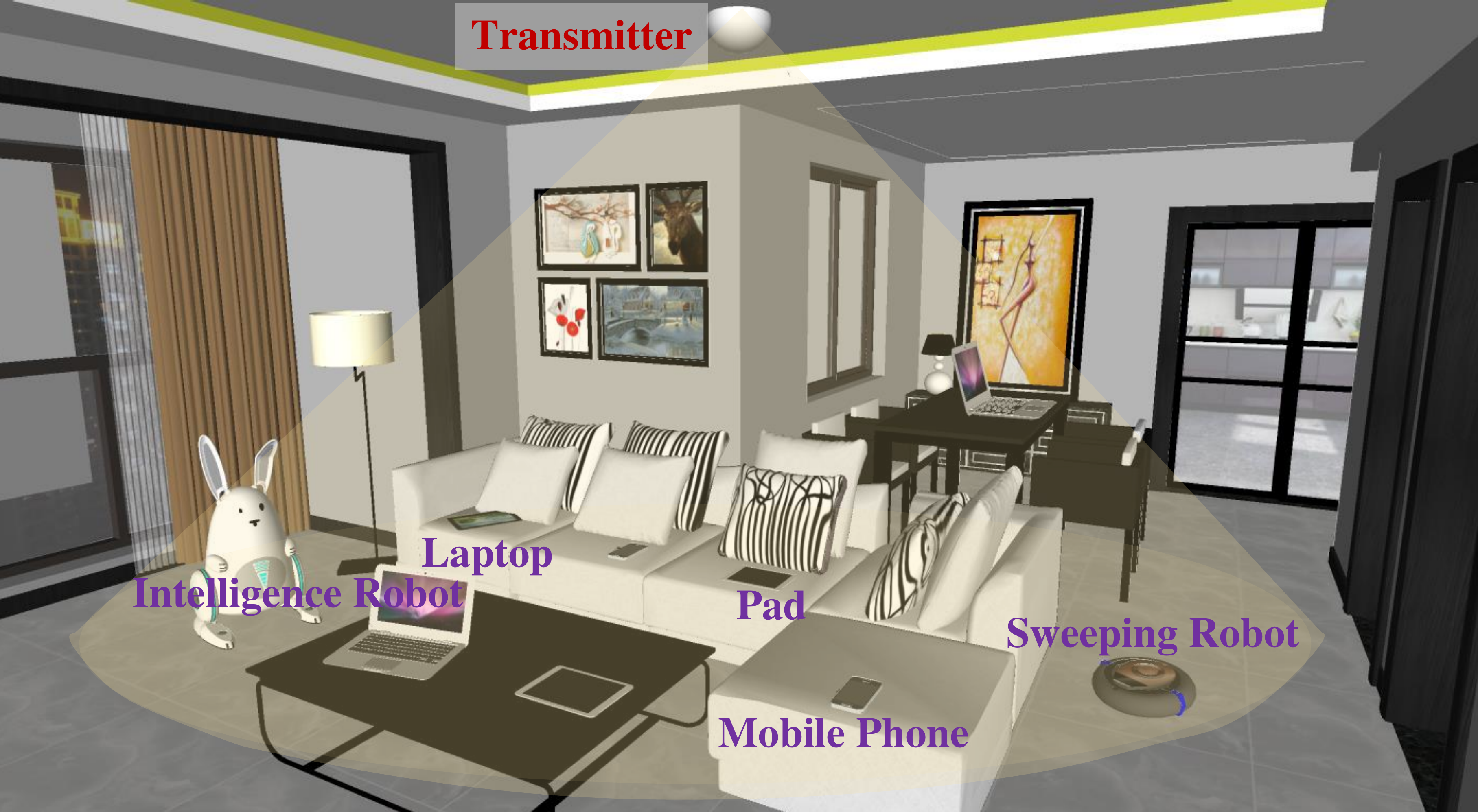}
	\caption{RBC mobile application scenario.}
    \label{Fig1Scenario}
\end{figure}

Similar to cellular base stations, the RBC transmitters placed at different locations have different charging coverage (the area within which the charging power received by mobile devices is larger than a certain threshold) \cite{fov}. Even within the coverage, it is very often to encounter the settings where multiple mobile devices, e.g., mobile phone, intelligent logistics robots and reconnaissance Unmanned Aerial Vehicles (UAV), require wireless charging service simultaneously \cite{zeng2019energy}. Similar to wireless communication systems, there are energy losses in the wireless energy transmission channel, so each receiver's received power is different when placed at or moving to different positions \cite{wang2018channel}. Figure \ref{Fig1Scenario} presents an RBC mobile application scenario: an RBC transmitter is placed at the ceiling of a drawing room and multiple mobile devices including mobile phone, laptop, pad, sweeping robot, and intelligent robot are embedded with RBC receivers, each of which requires wireless charging services. These mobile devices may be moving with a random speed in a random direction, and their received charging power varies due to their different distances to the transmitter.

Consider a practical setting where multiple collaborative mobile receivers request charging at the same time but the transmitter's output power cannot meet all the charging requests. To extend the battery life of all receivers, scheduling algorithms are well-motivated to schedule the charging power, duration, and order in this context \cite{yu2013general, cheng2015d2d}. Existing scheduling algorithms for RBC system are all based solely on receivers' status. For example, in the First Access First Charge (FAFC) scheduling algorithm, the receiver that first access to the system is charged first, while the receiver with high priority (related to the receivers' remaining energy and the preferred charging power) is first charged in High Priority Charge (HPC) scheduling algorithm \cite{fang2018, fang2018earning}. Since the receivers' charging power depends critically on the transmission channel relative to the transmitter, an effective charging scheduling algorithm should also account for the RBC energy transmission channel.

In a nutshell, the contributions of this paper include:

C1) A Channel-Dependent Charge (CDC) scheduling algorithm for controlling the charging power, duration, and order of mobile devices is put forth to extend the battery life of all receivers;

C2) To implement the CDC scheduling algorithm, the RBC transmitter's coverage, the receiver dynamic moving model, and related parameters including the receivers' initial locations and charging time slots are analyzed; and,

C3)  The performance of the CDC algorithm is numerically tested, and the results include: i) Compared with the Round-Robin Charge (RRC) scheduling algorithm, the performance of the CDC is superior, and ii) the measures to improve the CDC performance are prolonging the charging duration, increasing the input electric power, or limiting the number of receivers charging simultaneously within the coverage.



In the remainder of this paper, the RBC structure and transmission channel are described in Section II. Section III presents the scheduling principles behind and develops the CDC scheduling algorithm. Section IV discusses the mathematical model of the CDC algorithm. Performance of the CDC algorithm is evaluated in Section V, while this paper is concluded along with directions for future research.

\section{Transmission Channel Model}\label{Section2}

This section outlines the structure, the energy transfer process, and the energy transmission channel model of an RBC system.

\subsection{Resonant Beam Charging system}

\begin{figure}[!t]
	\centering
    \includegraphics[scale=0.48]{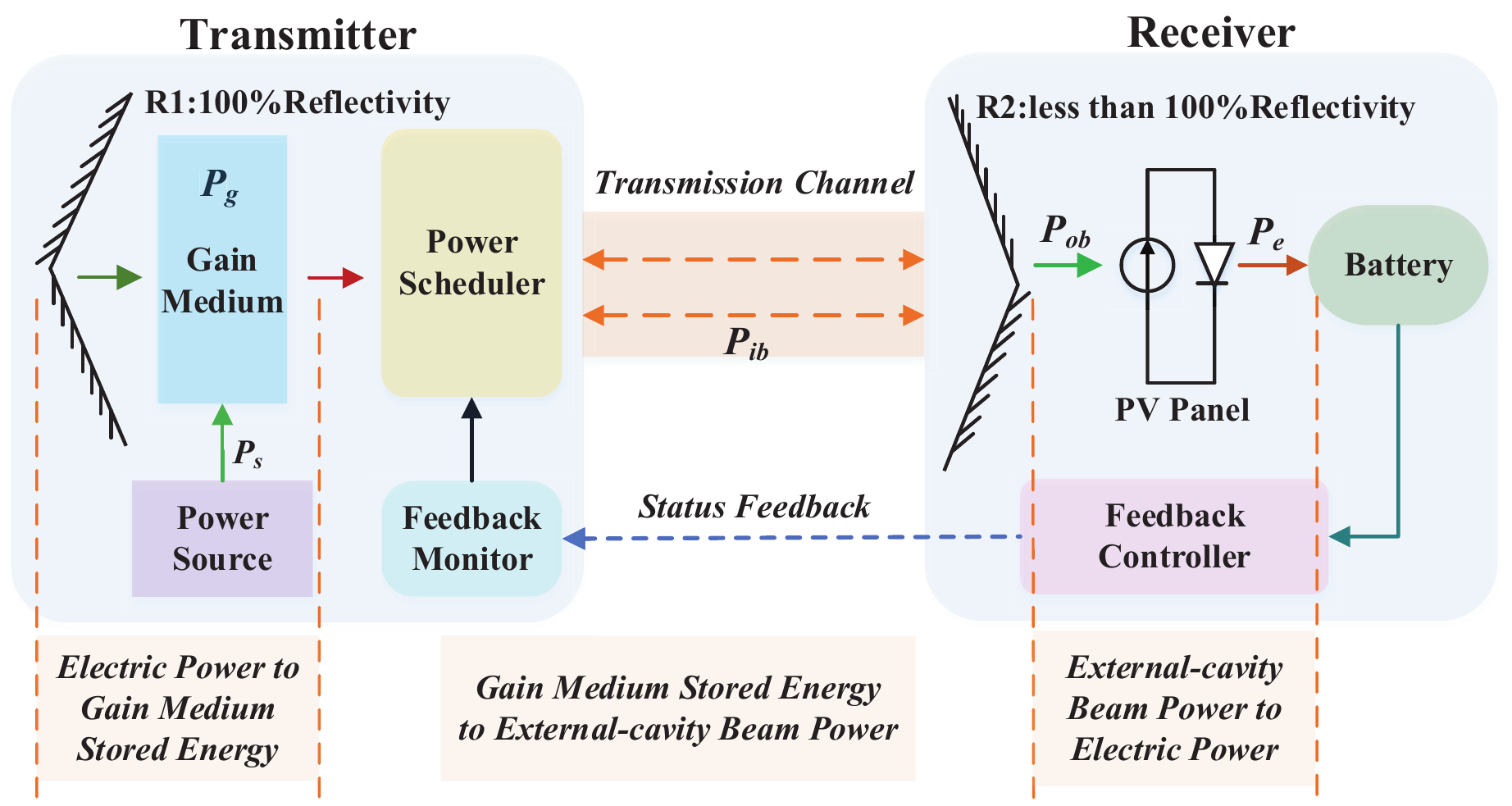}
	\caption{The RBC system structure.}
    \label{Fig1}
    \end{figure}

The structure of an RBC system is shown in Fig. \ref{Fig1}, which comprises a transmitter and a receiver. Specifically, the RBC transmitter includes a power source, a retro-reflector with $100$\% reflectivity (R1), a gain medium, a power controller, and a feedback monitor. The receiver consists of a retro-reflector with less than $100$\% reflectivity (R2), a Photovoltaic (PV) panel, a battery, and a feedback controller. The interspace between R1 and R2 is the resonant cavity \cite{fang2018earning}.
The battery in the receiver can be charged when the RBC system is on, and this charging process can be described in the following stages:

S1) The gain medium is pumped with the input electric power $P_s$;

S2) The power scheduler determines the charging order, power, and duration of receivers based on the status information collected by the feedback monitor;

S3) The scheduled beam $P_{ib}$ is transmitted to R2 and passes through R2 in part to form the receiver output beam;

S4) The beam power $P_{ob}$ is converted to the output electric power $P_e$ by a PV panel for battery charging \cite{aziz2014simulation}; and,

S5) The feedback controller collects the receiver's real-time information including the position, the remaining energy, and the charging time, and feeds them back to the feedback monitor in the transmitter. The method of feeding back information can be an existing communication method such as Wi-Fi or Bluetooth, or a novel communication approach through the resonant beam \cite{xiong2019resonant}.

To install an RBC system, the RBC transmitter can be manufactured as a transmitting device similar to the router, and installed at the top of the room or the car. The RBC receiver can be embedded in mobile devices that require wireless charging services, e.g., mobile phones, and tablets. The embedding methods include: i) the receiver is embedded into the new devices as an internal component, and ii) make the receiver as an external plug-in and connect it to the pre-existing devices through the Universal Serial Bus (USB) interface \cite{fov}.

\subsection{Wireless energy transmission channel model}

From Fig. \ref{Fig1} and \cite{koechner2013solid, penzkofer1988solid}, the battery's charging process in an RBC system can also be divided into three stages: i) the input electric power $P_s$ to the gain medium stored power $P_g$, in which the gain medium can be accumulated with the power provided by the power source; and, ii) the gain medium stored power $P_g$ to external-cavity beam power $P_{ob}$. The stored power is pumped and converted to intra-cavity beam power $P_{ib}$, which is subsequently transmitted to the receiver $R2$ to form the external-cavity beam power $P_{ob}$; and, iii) the external-cavity beam power $P_{ob}$ to output electric power $P_e$ by the PV panel. The power parameters corresponding to symbols are recorded in Table \ref{power}. The RBC transmission channel model is formulated on this basis, and the transmission stages of the channel are elaborated below.

\begin{table}[!t]
    \setlength{\abovecaptionskip}{0pt}
    \setlength{\belowcaptionskip}{-3pt}
    \centering
        \caption{Power Parameters}
        \vskip .05in
    \begin{tabular}{C{1.5cm} C{4.5cm}}
    \hline
     \textbf{Symbol} & \textbf{Parameter}\\
    \hline
    \bfseries{$P_s$} & {Input electric power}\\
    \bfseries{$P_g$} & {Stored power in gain medium }\\
    \bfseries{$P_{ob}$} & {External-cavity beam power}\\
    \bfseries{$P_{ib}$} & {Intra-cavity beam power}\\
    \bfseries{$P_e$} & {Output electric power}\\
    \hline
    \label{power}
    \end{tabular}
    \end{table}


The relationship between the input electric power $P_s$ and the gain medium stored power $P_g$ can be depicted as \cite{koechner2013solid, Qing2017}
\begin{equation}\label{eq1}
    P_{g} = \eta_g P_s
\end{equation}
where $\eta_g$ is the conversion efficiency of the input electric power to the stored power.


On the other hand, the gain medium stored power can be converted into the intra-cavity beam power by the spontaneous radiation \cite{koechner2013solid}. Then, the intra-cavity beam power transmits in the resonant cavity between R1 and R2, and passes through R2 partially to form the external-cavity beam \cite{liu2016dlc}. Thus, the gain medium stored power can be converted into the external-cavity beam power, and the conversion efficiency depends on transmission distance and efficiency \cite{zhang2018distributed2}.

The key factor influencing transmission efficiency is the diffraction loss. In the resonant cavity, the single transmission diffraction loss $\delta$ of the resonant beam can be presented as
\begin{equation}\label{eq2}
    \delta = e^{-2 \pi \frac{a^2}{\lambda (l + d)}}
\end{equation}
where $a$ is the radius of the retro-reflector R1 and R2, $l$ is the distance between the gain medium and R1, $d$ is the distance between the gain medium and R2, and $\lambda$ is the wavelength of the resonant beam.

The relationship between the gain medium stored power $P_{g}$ and the external-cavity beam power $P_{ob}$ can be depicted as \cite{koechner2013solid}
\begin{equation}\label{eq3}
    P_{ob} = \alpha P_g + C
\end{equation}
where $C$ is a constant related to the RBC system internal parameters, and $\alpha$ is the function relying on the transmission distance $d$ and the diffraction loss $\delta$
\begin{equation}\label{eq4}
    \alpha = \frac{2 (1 - f) m}{(1 + f) \delta - (1 + f) \ln f}
\end{equation}
where $f$ is the reflectivity of the output mirror R2, and $m$ is the overlap efficiency \cite{wang2018channel}.


Finally, the external-cavity beam power can be converted into the output electric power by the PV panel, whose function is similar to the solar panel. As presented in \cite{Qing2017}, when the PV panel works at the Maximum Power Point (MPP), the conversion efficiency of the beam power to electric power is maximum. The relationship between the external-cavity beam power $P_{ob}$ and the electric power $P_e$ can be stated as
\begin{equation}\label{eq5}
    P_{e} = \beta P_{ob} + \gamma
\end{equation}
where $\beta$ and $\gamma$ are constant coefficients, which can be obtained from \cite{Qing2017}.


Based on the analysis results of the above three stages and \eqref{eq1}-\eqref{eq5}, the relationship between the input electric power $P_s$ and the output electric power $P_e$ can be obtained as
\begin{dmath}\label{eq6}
    P_{e} = \beta (\alpha P_g + C) + \gamma \\
    = \beta \!\left[\!\left(\frac{2 (1 - f) m}{(1 + f)e^{-2 \pi \frac{a^2}{\lambda (l + d)}} - (1 + f) \ln f}\!\right) \eta_g P_s + C\!\right] + \gamma.
\end{dmath}

From \eqref{eq6}, the output electric power $P_e$ depends on the input electric power $P_s$ and the distance $d$ between the transmitter and the receiver. Moreover, due to the diffraction loss, the receiver's output electric power decreases as the distance $d$ increases for a constant input power $P_s$.

For a mobile device, the distance to the transmitter is decided by its location and movement. The closer the receiver to the transmitter is, the higher charging power it gets. Thus, when multiple mobile devices request charging simultaneously, the receiver closer to the transmitter acquires more energy, at a higher charging efficiency than receivers far away.

Building the channel model and analysis above, the CDC scheduling algorithm will be developed in the next section to extend the receivers' battery life.

\section{Scheduling Design}\label{Section2}

To design the CDC scheduling algorithm, the scheduling principle and the execution flow will be presented in detail in this section.

\subsection{Scheduling principle}\label{}
In the RBC mobile application scenario, multiple mobile receivers, which request charging simultaneously, are at different locations and may even be moving, so their distances to the transmitter vary over time. Using \eqref{eq6}, the charging power received by per receiver is different. The shorter the distance, the higher charging power receiver can get. To extent the battery life of all receivers, the charging sequence of the receivers is determined by considering the two factors of the receivers' battery remaining energy and distance to the transmitter.

A scheduling coefficient $F_{s}$ related to the remaining energy $E_{r}$ and the distance to the transmitter $d$ is set for each receiver requesting for charging. In the meanwhile, all receivers that request charging are arranged into a charging queue from the smallest scheduling coefficient to the largest, and the receiver with the minimum coefficient is first charged at each time slot. The scheduling coefficient $F_{si}$ of the receiver $R_i$ can be obtained by
\begin{equation}\label{eqcoefficient}
    F_{si} = C_e E_{ri} + C_d d_{i}, \ \  E_{ri} \in [0,  E_b], d_i \in [0, D]
\end{equation}
where $E_{ri}$ is remaining energy and $d_{i}$ is the distance to the transmitter of the receiver $R_i$. $C_e$ and $C_d$ are the constant coefficient. In the CDC algorithm, two factors of the battery remaining energy and the distance are equally important. Thus, $C_e$ and $C_d$ are both $0.5$. In addition, $E_b$ is the battery total energy, and $D$ is the longest distance between the transmitter and the receiver within the transmitter's coverage.

The scheduling strategy means that the receiver with the minimum remaining energy and the shortest distance to the transmitter can be first charged with the maximum charging power. Moreover, the receiver closer to the transmitter will be earlier charged between the two receivers with the same remaining energy. This receiver can get a relatively high charging power based on \eqref{eq6}, so the higher charging efficiency can be achieved.


\begin{figure}[!t]
    \centering
    \includegraphics[scale=0.6]{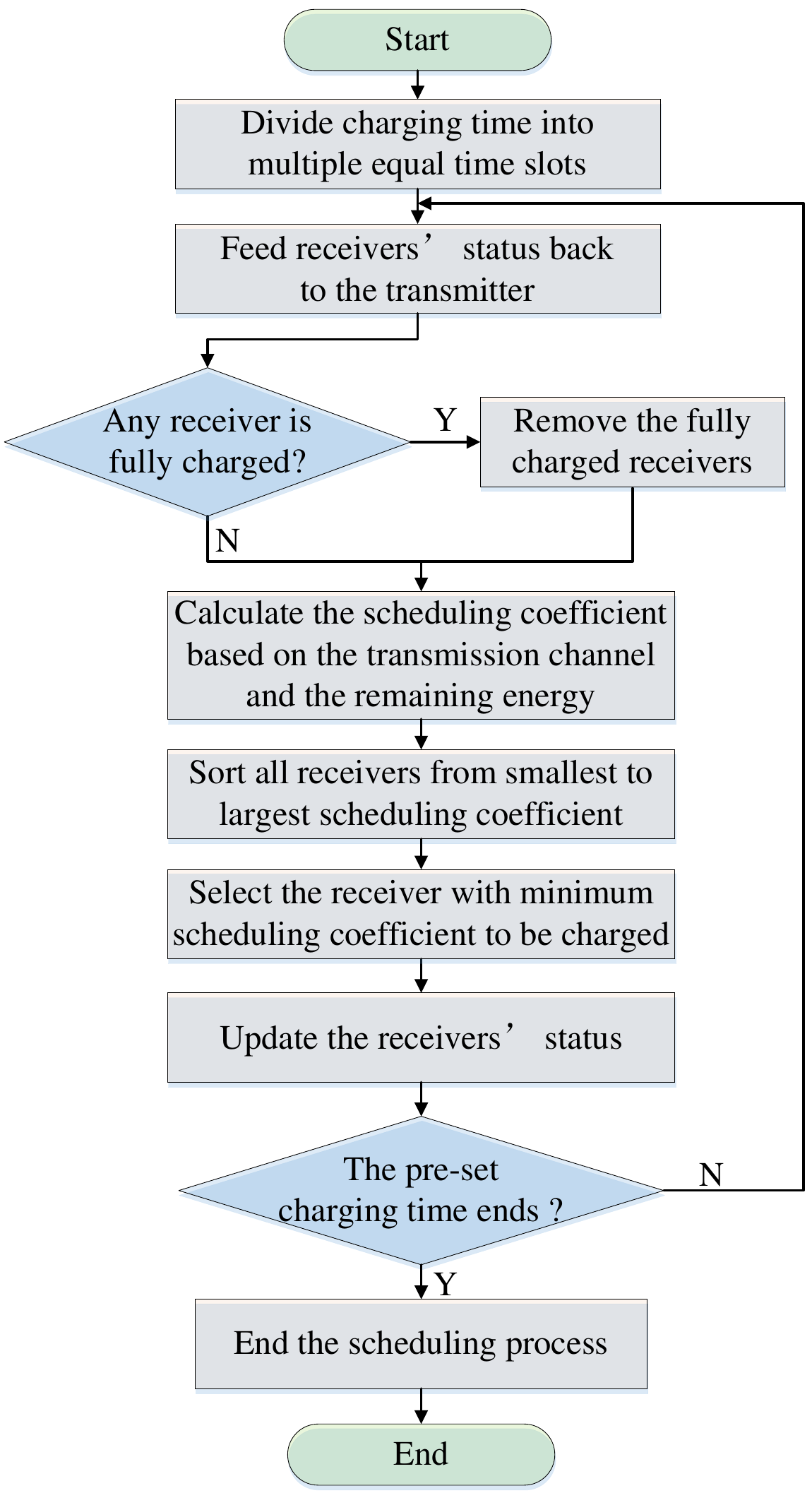}
	\caption{Execution flow of the CDC algorithm.}
    \label{Fig2}
\end{figure}


\subsection{Execution flow}\label{}

In the RBC mobile application scenario, the receivers with different remaining energy may be moving randomly, and thus the transmission channels between the RBC transmitter and receivers may vary constantly. Relying on the scheduling principle, the execution flow of the CDC algorithm is depicted in Fig. \ref{Fig2}. The detailed steps are described below:

S1) The charging duration is divided into multiple equal unit-length charging time slots;

S2) The receivers' feedback controllers monitor their status (e.g., the location, the remaining energy, charging duration, to name a few) and feed them back to the transmitter through the information loaded by resonant beam, Wi-Fi or bluetooth;

S3) Detect the presence of a fully charged receiver. If so, remove the receiver from the queue to be charged;

S4) Calculate the scheduling coefficient $F_{s}$ for each receiver based on the transmission channel and remaining energy;

S5) All receivers are sorted from the smallest scheduling coefficient to the largest;

S6) The receiver with the minimum scheduling coefficient is selected to be charged with power calculated in \eqref{eq6};

S7) At the end of a charging time slot, the receivers' status is updated; and,

S8) If the pre-set charging duration is achieved, the scheduling process terminates. Otherwise, turn to S2), and start a new charging time slot.

With the above CDC scheduling process, the receivers within the transmitter's coverage will be optimally scheduled charging to extend the battery life of all receivers.


\section{Scheduling Model}
To implement the CDC scheduling algorithm, the RBC transmission channel model, the transmitter's coverage, and the receivers' dynamic moving model will be analyzed. In addition, the executable pseudo code will be described.

    \begin{table}[!t]
    \setlength{\abovecaptionskip}{0pt}
    \setlength{\belowcaptionskip}{-3pt}
    \centering
        \caption{Transmission Channel Model Parameters}
        \vskip .05in
    \begin{tabular}{C{1.5cm} C{3cm} C{1.8cm}}
    \hline
     \textbf{Symbol} & \textbf{Parameter} & \textbf{Value}  \\
    \hline
    \bfseries{$\beta$} & {Constant coefficient} & {0.3487} \\
    \bfseries{$f$} & {R2 reflectivity} & {88\%} \\
    \bfseries{$m$} & {Overlap efficiency} & {80\%} \\
    \bfseries{$\pi$} & {Circular constant} & {3.14} \\
    \bfseries{$a$} & { R1 and R2 radius} & {1.5mm} \\
    \bfseries{$\lambda$} & {Beam wavelength} & {1.064e-06m} \\
    \bfseries{$\eta_g$} & {Conversion efficiency} & {28.49\%} \\
    \bfseries{$C$} & {Constant coefficient} & {-5.64} \\
    \bfseries{$\gamma$} & {Constant coefficient} & {-1.535} \\
    \hline
    \label{channel}
    \end{tabular}
    \end{table}
\subsection{Transmission channel model quantification}\label{}

\begin{figure}[!t]
    \centering
    \includegraphics[scale=0.6]{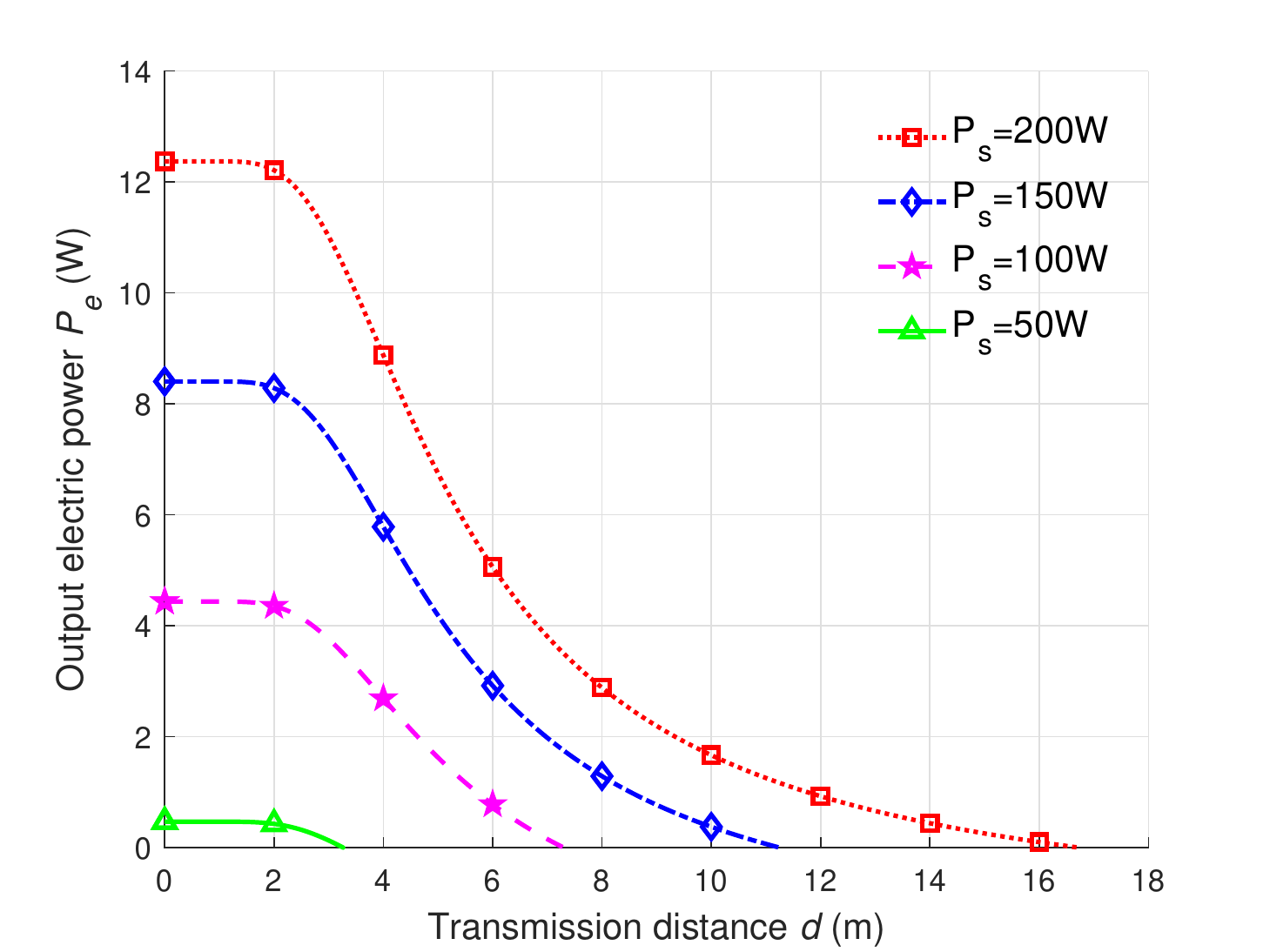}
	\caption{Output electric power versus transmission distance.}
    \label{Fig3}
\end{figure}

The parameters (e.g., $\beta, R, m, \pi$) for \eqref{eq6} are specified in Table \ref{channel} \cite{koechner2013solid, Qing2017}. The output electric power $P_e$ is plotted as a function of the transmission distance $d$ with the input electric power $P_s=50$, $100$, $150$, and $200$W, respectively in Fig. \ref{Fig3}.

From Fig. \ref{Fig3}, when the transmission distance $d$ is small ($<=2m$), the output electric power is almost constant, but gradually decreases to $0$ as $d$ increases from $2m$. For a fixed distance $d$, the higher the input electric power $P_s$, the larger the output power $P_e$. The largest transmission distance $D$ (the output electric power is $0$) increases with $P_s$ and the values of $D$ over different $P_s$ are shown in Table \ref{largest}. For example, the largest transmission distance is $7.2972m$ when the input electric power is $100W$.

\begin{table}[!t]
    \setlength{\abovecaptionskip}{0pt}
    \setlength{\belowcaptionskip}{-3pt}
    \centering
        \caption{Largest Transmission Distance}
        \vskip .05in
    \begin{tabular}{C{3.5cm} C{4.8cm}}
    \hline
     \textbf{Input Electric Power(W)} & \textbf{Largest Transmission Distance(m)}  \\
    \hline
    \bfseries{50} & {3.2623} \\
    \bfseries{100} & {7.2972} \\
    \bfseries{150} & {11.2438} \\
    \bfseries{200} & {16.7149} \\
    \hline
    \label{largest}
    \end{tabular}
    \end{table}


\subsection{Transmitter coverage}\label{}
\begin{figure}[!t]
    \centering
    \includegraphics[scale=0.6]{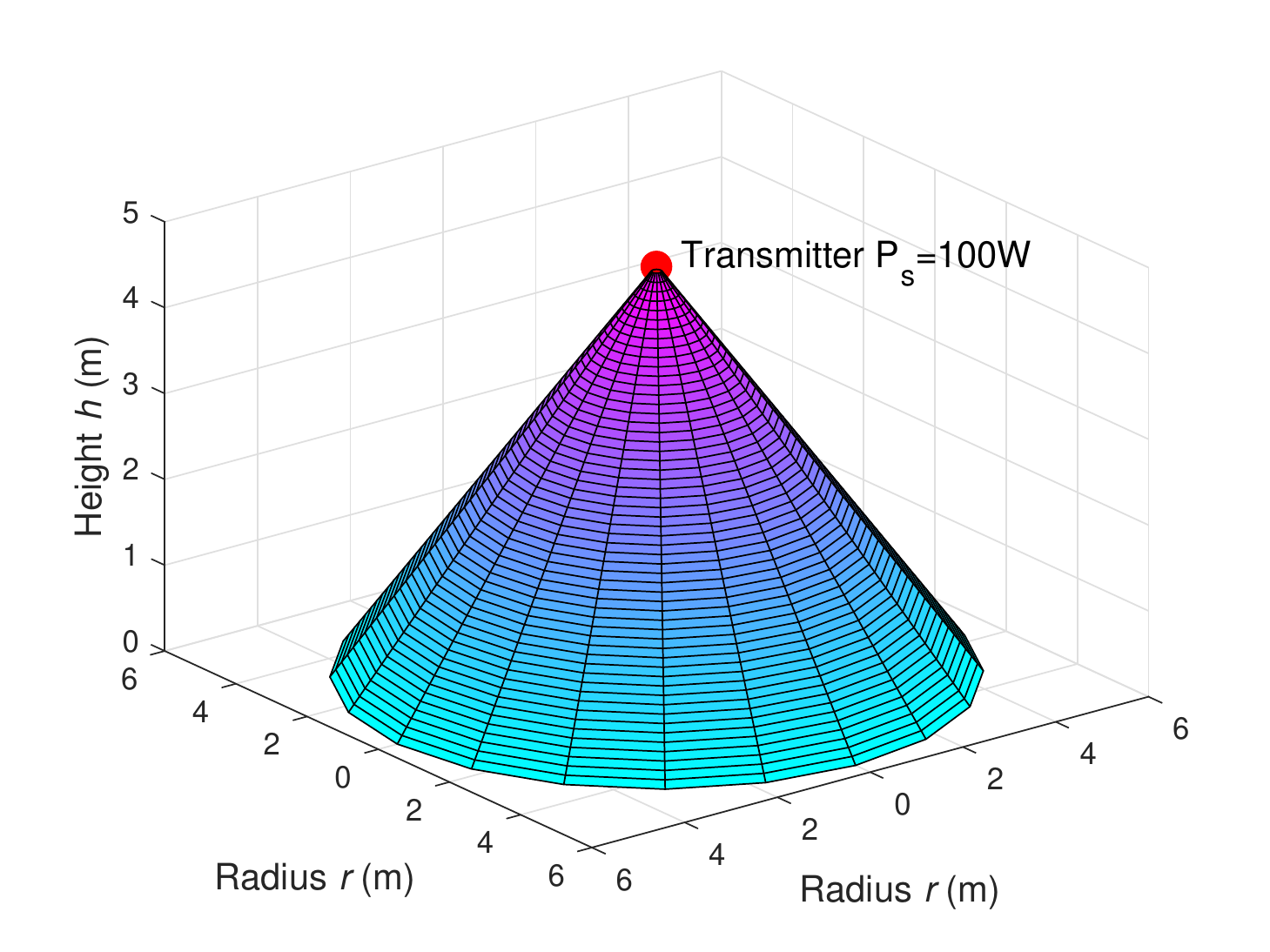}
	\caption{RBC Transmitter coverage ($P_s=$100W).}
    \label{Fig4}
\end{figure}
The RBC transmitter can be installed in the ceiling to provide wireless charging service for mobile devices in the coverage. An RBC transmitter's coverage is decided by the Field of View (FOV) $T_{fov}$ and the largest transmission distance $D$. The FOV of an RBC system is set as $100^{\circ}$ (i.e., $T_{fov} = 100^{\circ}$) according to the technical product of Wi-Charge \cite{fov}, while the largest transmission distance $D$ can be decided by the input electric power.

To compute the coverage of an RBC transmitter, the height $h$, the radius $r$ of the basal plane, and the largest transmission distance $D$ constitute a right triangle. The relationship between $h$ and $D$ is thus
\begin{equation}\label{eq7}
    h = D \times \cos\!\left(\frac{T_{fov}}{2}\!\right)
\end{equation}
and the radius $r$ is
\begin{equation}\label{eq8}
    r = D \times \sin\!\left(\frac{T_{fov}}{2}\!\right).
\end{equation}

From \eqref{eq6}, \eqref{eq7}, and \eqref{eq8}, the transmitter coverage relies on the input electric power $P_s$. When $P_s$ is $100$W, the corresponding transmitter coverage is depicted in Fig. \ref{Fig4}. A transmitter's coverage is a three-dimensional region forming a cone. In addition, the shape of the transmitter's coverage with different $P_s$ is the same, but their height $h$ and radius $r$ are diverse. The height and radius of $P_s = $ $50$W, $100$W, $150$W and $200$W transmitter's coverage are shown in Table \ref{coverage}. From Table \ref{coverage}, given the FOV, the higher $P_s$, the larger $h$ and $r$. The vertical height between the transmitter and receiver and the radius of the coverage area are $4.6906m$ and $5.5900m$ respectively when the input electric power is $100W$. Thus, the coverage increases as the input electric power grows.

\begin{table}[!t]
    \setlength{\abovecaptionskip}{0pt}
    \setlength{\belowcaptionskip}{-3pt}
    \centering
        \caption{Transmitter Coverage Parameters}
        \vskip .05in
    \begin{tabular}{C{3.5cm} C{1.5cm} C{1.8cm}}
    \hline
     \textbf{Input electric Power (W)} & \textbf{Height (m)} & \textbf{Radius (m)}\\
    \hline
    \bfseries{50 } & {2.0970} & {2.4991} \\
    \bfseries{100 } & {4.6906} & {5.5900} \\
    \bfseries{150 } & {7.2274} & {8.6133} \\
    \bfseries{200 } & {10.7441} & {12.8044} \\
    \hline
    \label{coverage}
    \end{tabular}
    \end{table}

\begin{figure}[!t]
    \centering
    \includegraphics[scale=0.6]{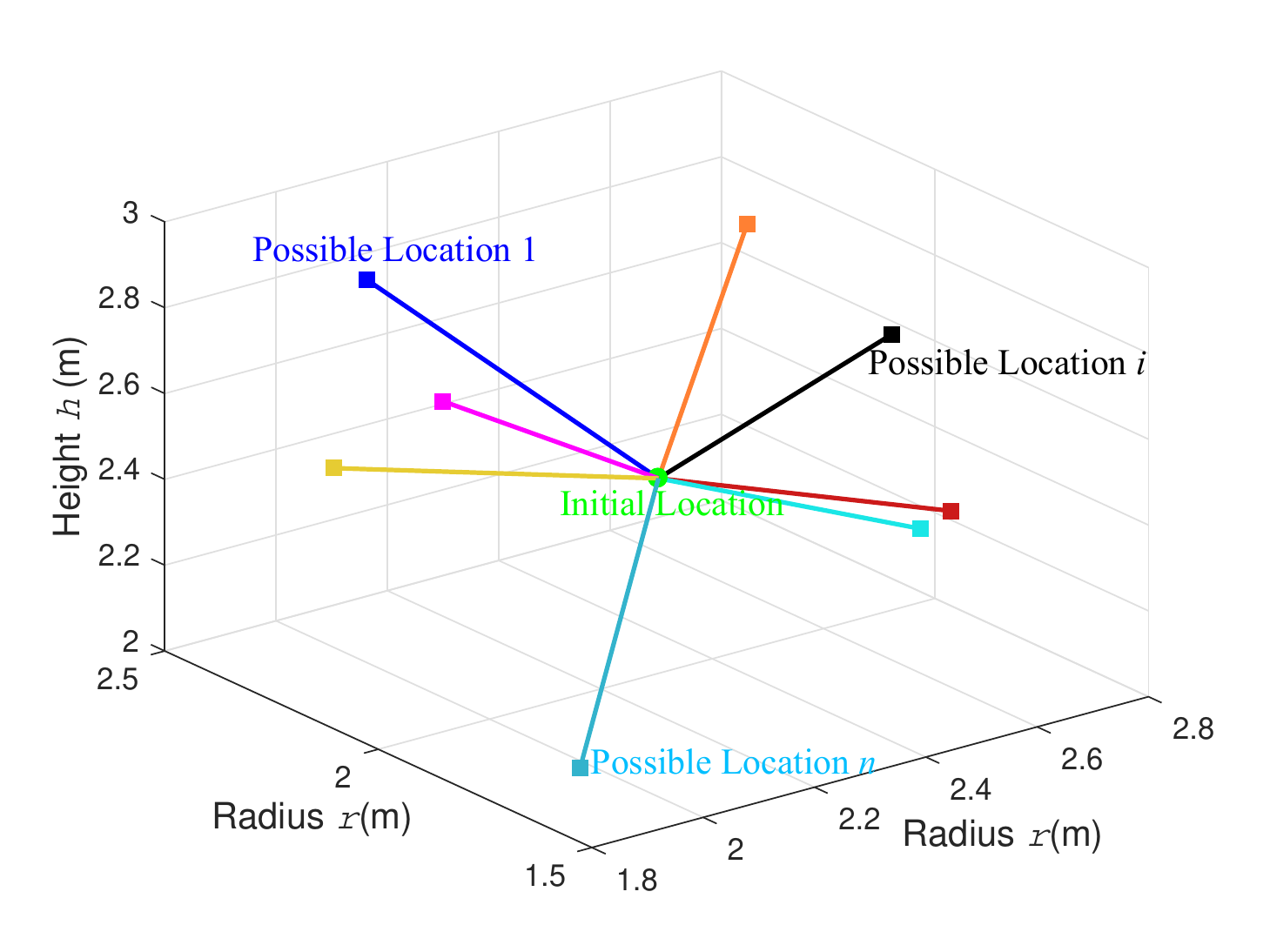}
	\caption{A receiver dynamic moving example.}
    \label{Fig3move}
\end{figure}
\subsection{Receiver dynamic moving model}\label{}

During the charging process, receivers within the transmitter's coverage may be moving at random directions with random velocities. For example, a mobile phone can be picked up, moving at any speed with a person, and it can also be put down anytime. Set the transmitter's installation position as the origin of Cartesian Coordinate System, the receivers' dynamic moving model is described below:

1) To initialize a charging process, the location of the receiver $R_i$ is randomly set as $(x, y, z)$ based on the Cartesian Coordinate System of the transmitter's coverage.

2) At a charging time slot, suppose receiver $R_i$ moves at a constant velocity $v$ in random directions.

3) The velocity $v$ of $R_i$ can be split into its components in the $x$, $y$, and $z$ directions as $v_x$, $v_y$, and $v_z$.

4) After $R_i$ moves for a period of time $t$, the location of $R_i$ turns into $(x + v_x t, y + v_y t, z + v_z t)$. Thus, the distance between $R_i$ and the transmitter is
\begin{equation}\label{eq9}
    d = \sqrt{(x + v_x t)^2 + (y + v_y t)^2 + (z + v_z t)^2}.
\end{equation}

5) Finally, the charging power for $R_i$ can be calculated based on the wireless energy transmission channel model.

The moving velocity of the receiver is random, depending on the actual situation of the receiver. For example, in this paper, the velocity in $x, y$ and $z$ directions is assumed as the value between $0$ and $0.01$m/s (i.e., $v_{max}=0.01$m/s). In Fig. \ref{Fig3move}, an example of the receiver dynamic moving model is presented, where the receiver's initial location is the green dot. Then, the squares are used to mark eight possible locations after the receiver moved for some time, and the lines between the dot and squares are the straight lines between the initial location and the possible locations. For instance, the device may move to the possible locations $1$, $i$ or $n$ after moving for a spell.

\subsection{Executable pseudo code}\label{}

In the CDC algorithm, the related parameters are elaborated as follows:

1) The number of receivers within the RBC transmitter's coverage $N_r$ is random during a scheduling process.

2) The receivers' initial locations are random within a transmitter's coverage. Thus, the initial distances between the transmitter and the receivers are also random.

3) The percentages of the remaining capacity of all receivers' batteries are the random number between $0\%-100\%$, and the remaining energy is the product of the percentage of the remaining capacity and the total energy of the battery $E_b$.

4) The duration of a charging time slot $T_c$ depends on the total charging duration and the charging status (e.g. the number of receivers, the input electric power, to name a few). Moreover, the charging power received by a receiver is constant per time slot. It is equal to the power received at the beginning of the time slot.

\begin{algorithm}[!h]
    \caption{CDC Algorithm}
    \begin{algorithmic}[1]
    \Require $N_r$;
    \State initialize $T_{c}$, $T_{o}$, $\beta$, $f$, $m$, $\pi, a, \lambda, \eta_g, C, \gamma$, $P_s$, $E_b$;
    \State $T_{s} \leftarrow 0$;
    \State $P_e \leftarrow 0$, using \eqref{eq6} calculate $D$;
    \State $h \leftarrow D \times cos (\frac{T_{fov}}{2})$;
    \State $r \leftarrow D \times sin (\frac{T_{fov}}{2})$;
    \State $R(:,1) \leftarrow rand \times (h-0)$;
    \State $R(:,2) \leftarrow rand \times [(h-R(:,1)) \times tan\big(\frac{T_{fov}}{2}\big)-0]$;
    \State $R(:,3) \leftarrow rand \times [sqrt(r^2 - R(:,2)^2)-0]$;
    \State $R(:,4) \leftarrow \frac{rand \times [100 - 0]}{100} \times E_b$;
    \While {$ T_s \leqslant T_o$}
    \State initialize $P_s$;
    \State $R(:,5) \leftarrow \sqrt{R(:,1)^2 + R(:,2)^2 + R(:,3)^2}$;
    \State $R(:,6) \leftarrow C_e \times R(:,4) + C_d \times R(:,5)$;
    \State $R_{sort} \leftarrow sortrows(R, 6)$;
    \State $d \leftarrow R_{sort}(1, 5)$;
    \If {$d <= D \&\& R_{sort}(1, 4)<E_b$
    \State $P_{e} \!\leftarrow \!\beta \!\biggl[\!\biggl(\frac{2 (1 - f) m}{(1 + f)e^{-2 \pi \frac{a^2}{\lambda (l + d)}} - (1 + f) \ln f}\!\biggr) \eta_g P_s \!+\! C\!\biggr] + \gamma$;
    \State $R_{sort}(1, 4) \leftarrow R_{sort}(1, 4) + P_{e} \times T_c$;
    \EndIf
    \State Restore the sorting of the array R;}
    \State $T_s \leftarrow T_s + T_c$;
    \State $v(:,1|2|3) \leftarrow \pm rand \times (v_{max}-0)$;
    \State $R(:,1|2|3) \leftarrow R(:,1|2|3) + v(:,1|2|3) \times T_c$;
    \EndWhile
    \State $E_s \leftarrow sum(R(:,4))$;
    \State \Return{$E_s$};
    \label{HPC}
    \end{algorithmic}
    \end{algorithm}
Based on the scheduling principle and the quantitative parameters of the CDC algorithm, its executable pseudo code can be programmed as Algorithm 1. The core executable steps are depicted below:

S1) Initialize parameters including $T_{c}$, $T_{o}$ (the pre-set charging duration), $\beta$, $f$, $E_b$;

S2) The total charging duration $T_s$ is set to be $0$;

S3) When the output electric power $P_e$ is $0$ with constant input electric power, calculate the longest transmission distance $D$ using \eqref{eq6}. The height $h$ and radius $r$ can be computed by \eqref{eq7} and \eqref{eq8};

S4) The receiver initial height $R(:,1)$ (i.e. $z \in [0,h]$) is a random value within the height of coverage $h$, and radius $R(:,2)$ (i.e. $x\in[0, (h-R(:,1))\times \tan (\frac{T_{fov}}{2})]$) and $R(:,3)$ (i.e. $y \in [0, \sqrt{r^2 - R(:,2)^2}$]) are limited by $R(:,1)$ and the FOV;

S5) The remaining energy of receivers' battery $R(:,4)$ is set as the product of the random number between $0\%$ and $100\%$ and the total battery energy $E_b$;

S6) When the total charging duration $T_s$ is less than the pre-set charging duration $T_o$, the receivers are scheduled to be charged;

S7) Calculate the distances between all receivers and the transmitter $R(:, 5)$ and the scheduling coefficients $R(:, 6)$ of all receivers;

S8) Sort the receivers from smallest to largest scheduling coefficient. If the receiver with the minimum scheduling coefficient is within the chargeable coverage and not fully charged, it will be charged with the power computed by \eqref{eq6};

S9) Then, restore the sorting of the receiver array $R$. $T_s$ plus a time slot $T_c$, and the receivers may move with random velocities $V(:,1|2|3)$ at $x$, $y$, and $z$ directions. Thus, the receivers' locations $R(:,1|2|3)$ may change with the movements; turn to S6); and,

S10) When the scheduling process ends, calculate and return the sum of the remaining energy $E_s$ of all receivers.

Based on the above analysis, the scheduling principle and implementation of the CDC algorithm have been presented. However, the performance of the CDC algorithm and what factors may influence its performance need to be studied.

\section{Performance Analysis}

To evaluate the performance of the CDC algorithm, the performance of the CDC algorithm in terms of all receivers' average remaining energy under different influencing factors is analyzed. In addition, for proving the excellent charging performance, the performance of the CDC algorithm is compared with the Round-Robin Charge (RRC) scheduling algorithm under the same conditions.

In the performance analysis, the duration of a time slot is set to $10s$, and the total energy of the battery is set as $10.35Wh$ (the battery energy of iPhone X). The Monte Carlo realization is adopted to take multiple simulations with random variables, and the receiver' average remaining energy $E_{sa}$ is computed after multiple scheduling processes. $E_{sa}$ can be calculated by
\begin{equation}\label{eq10}
    E_{sa} = \frac{\sum\limits_{i=1}^{n} E_s}{N_r \times n}
\end{equation}
where $E_s$ is the total remaining energy of all receivers over a charging duration $T_s$. $N_r$ is the number of receivers, and $n$ is the number of simulations.

\subsection{Round-Robin Charge Algorithm}\label{}

To highlight the superiority of the CDC algorithm, the CDC algorithm is compared with RRC scheduling algorithm. The RRC algorithm is one of the most easily implemented scheduling algorithms in the RBC system. However, it is lack of consideration for the receiver's remaining energy and transmission channel. The implementation steps of the RRC algorithm include:

S1) The charging duration is divided into multiple equal unit-length charging time slots;

S2) The receivers are queued according to the time they connecting with the transmitter;

S3) During a charging time slot, the receiver in the head of the queue is selected to be charged with the power calculated by \eqref{eq6};

S4) At the end of a time slot, the receiver has been charged is assigned to the end of queue, and the charging duration plus a time slot; and,

S5) If the pre-set charging duration is achieved, the charging process ends, and the receiver's average charging energy can be obtained. Otherwise, turn to S3).

\subsection{Evaluation and Comparison}\label{}
\begin{figure}[!t]
    \centering
    \includegraphics[scale=0.6]{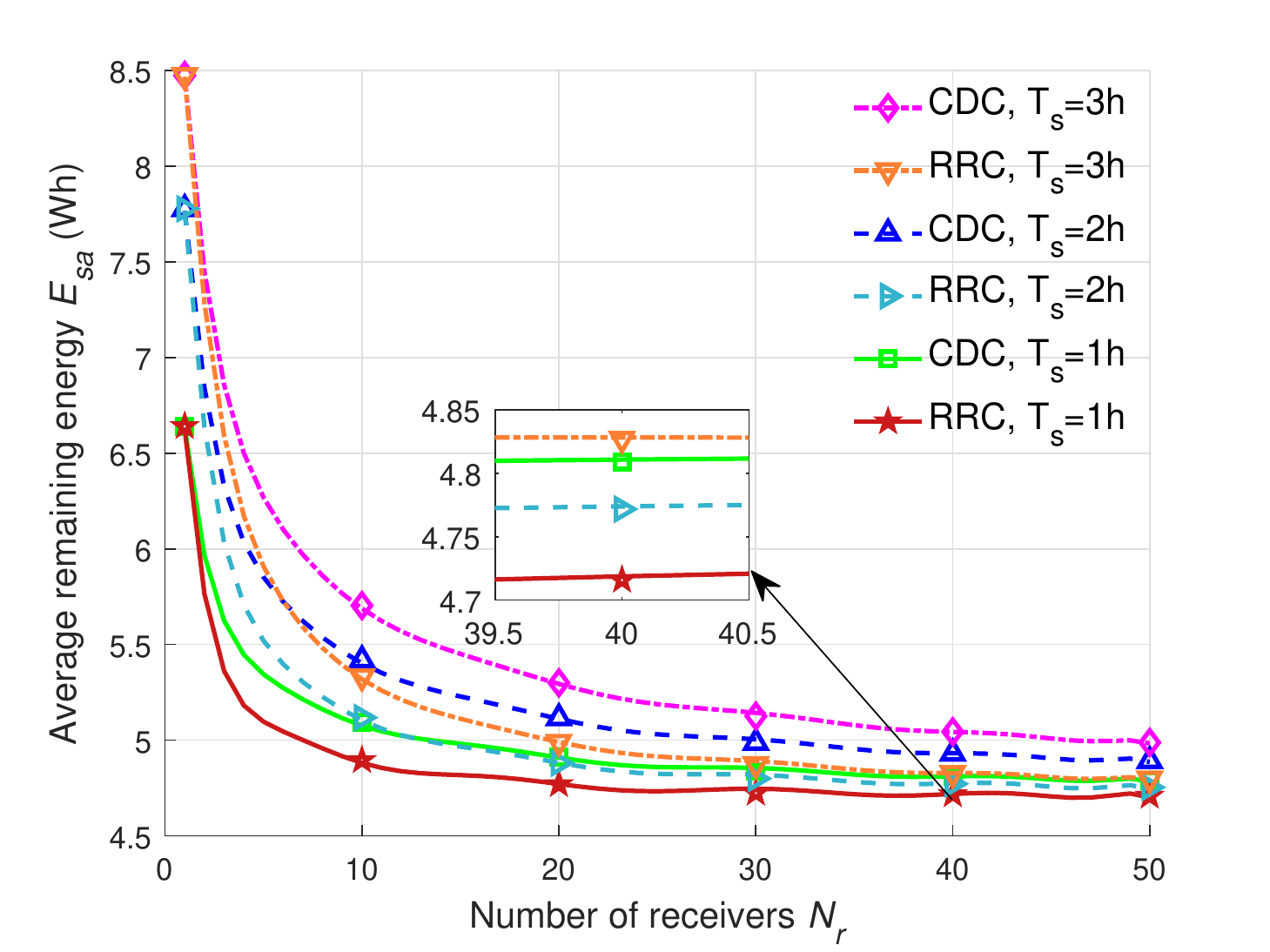}
	\caption{Average remaining energy versus the number of receivers (different charging duration).}
    \label{Fignumtime}
\end{figure}
The performance of the CDC algorithm can be affected by many factors, among which, the most direct impacts include: i) the number of receivers $N_r$, and, ii) the charging duration $T_s$, and, iii) the input electric power $P_s$. In this subsection, the performance of the CDC will be analyzed and compared with the RRC algorithm under these three influencing factors.

\noindent
\emph{E1) Different charging duration and number of receivers}

When the number of receivers $N_r$ within an RBC transmitter's coverage is between $1$ and $50$, the input electric power is $200$W, and the charging duration $T_s$ is $1$h, $2$h, and $3$h, the changes of the receivers' average remaining energy $E_{sa}$ over the number of receivers $N_r$ are presented in Fig. \ref{Fignumtime}.

For the CDC algorithm in Fig. \ref{Fignumtime}, since only one receiver with minimum scheduling coefficient can be charged during a time slot $T_c$, there exists a maximum charging energy in a charging duration. $E_{sa}$ decreases sharply when $N_r$ is small ($<=20$), but gradually decreases after $20$. Given the same number of receivers, $E_{sa}$ augments as charging duration prolongs.

For the RRC algorithm in Fig. \ref{Fignumtime}, similar to the CDC scheduling algorithm, $E_{sa}$ decreases with the increase of $N_r$ and increases with the increase of charging duration with the same $N_r$. What's more, when $N_r$ is greater than $1$, $E_{sa}$ of the RRC algorithm is less than that of the CDC algorithm with the same $N_r$ and $T_s$.

\begin{figure}[!t]
    \centering
    \includegraphics[scale=0.6]{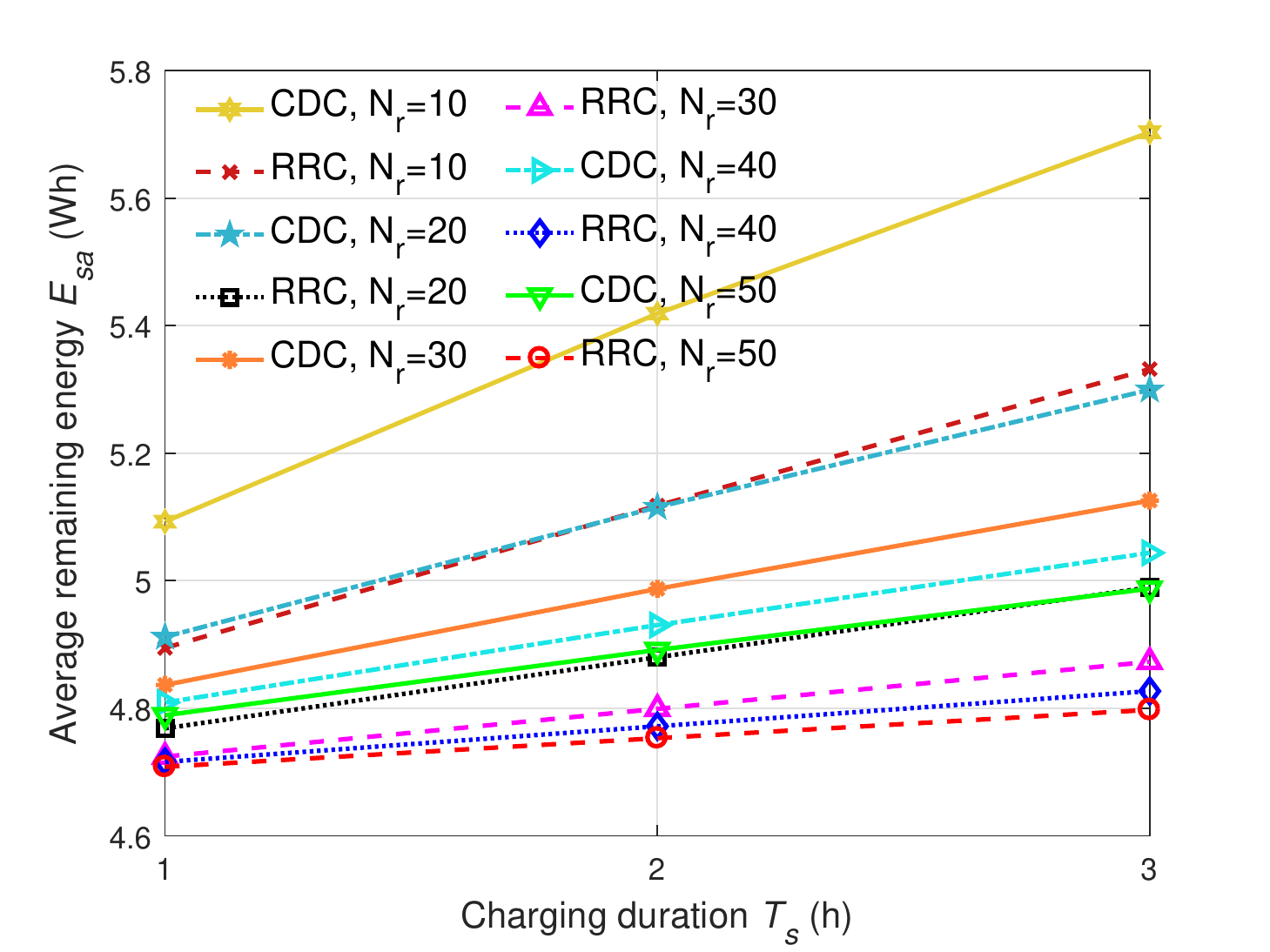}
	\caption{Average remaining energy versus charging duration (different number of receivers).}
    \label{Figtimenum}
\end{figure}

The changes of $E_{sa}$ over different charging durations $T_s$ are described in Fig. \ref{Figtimenum}.

In Fig. \ref{Figtimenum}, for CDC and RRC scheduling algorithms, $E_{sa}$ increases with the increase of charging duration despite what number of receivers is. For the same charging duration, $E_{sa}$ decreases as $N_r$ increases.
$E_{sa}$ of CDC algorithm is always higher than that of the RRC algorithm when $N_r$ is the same.
\begin{figure}[!t]
    \centering
    \includegraphics[scale=0.6]{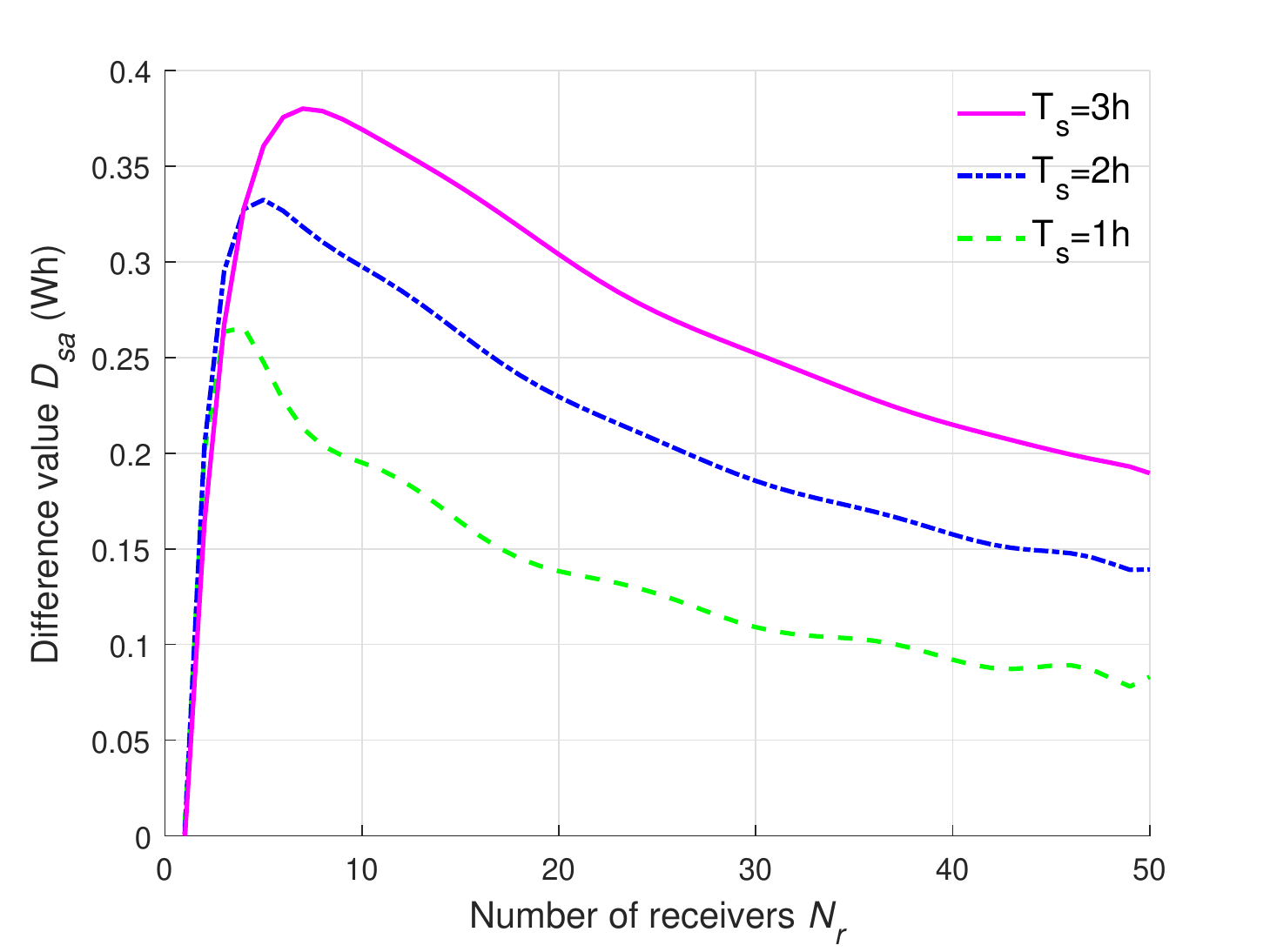}
	\caption{$E_{sa}$ difference value between CDC and RRC versus the number of receivers (different charging duration).}
    \label{differencetime}
\end{figure}

To illustrate the superiority of the CDC algorithm, $D_{sa}$ is defined as the difference value between the receivers' average remaining energy of the CDC algorithm and that of RRC algorithm. The changes in the difference $D_{sa}$ between $E_{sa}$ of the CDC algorithm and that of the RRC algorithm under different charging duration are depicted in Fig. \ref{differencetime}. $D_{sa}$ is equal to $0$ when $N_r$ is $1$. In addition, $D_{sa}$ increases rapidly first, then decreases gradually over $N_r$. When $N_r$ is same, $D_{sa}$ increases as the charging duration prolongs. Thus, compared with the RRC algorithm, the performance of the CDC algorithm is getting better with longer charging duration and a reasonable number of receivers (e.g. $N_r$ is from $2$ to $10$).

Thus, the performance of the CDC algorithm is better with the longer charging duration and the appropriate amount of receivers. Besides, the performance of CDC is better than that of RRC with the same input electric power, variable charging duration and the number of receivers.

\noindent
\emph{E2) Different input electric power and number of receivers}
\begin{figure}[!t]
    \centering
    \includegraphics[scale=0.6]{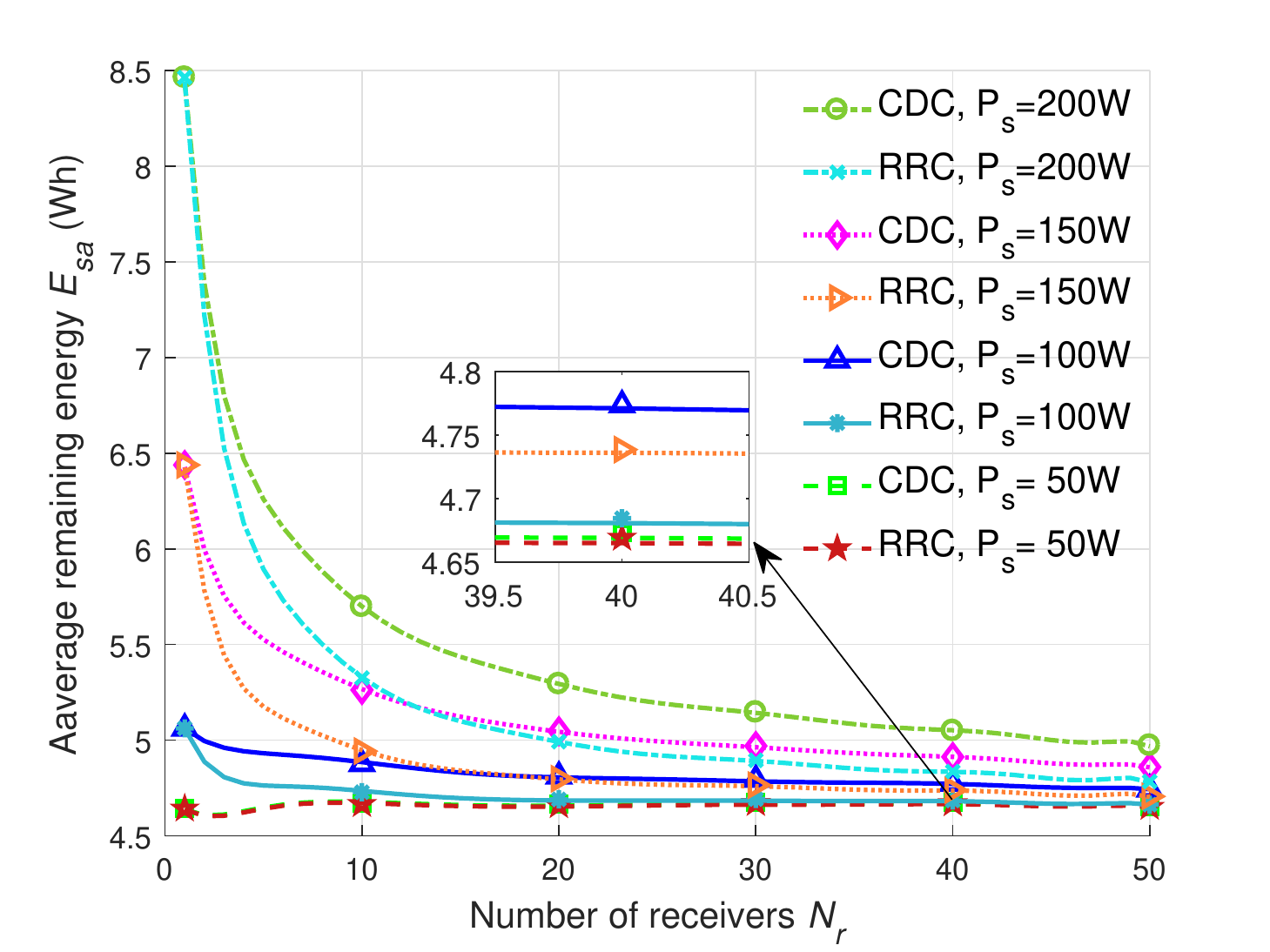}
	\caption{Average remaining energy versus the number of receivers (different input electric power).}
    \label{powernum}
\end{figure}
%

When the input electric power $P_s$ is $50$W, $100$W, $150$W, and $200$W, respectively, the transmitter's coverage is shown in Table \ref{coverage}, and the coverage of the transmitter with $200$W $P_s$ is largest. To guarantee the randomness of the initial locations of all receivers, whose number is from $1$ to $50$, their initial locations are limited within the coverage of the transmitter with $200$W input electric power.

After receivers are charged by the RBC transmitter with $50$W, $100$W, $150$W, and $200$W input electric power for $3$ hours, the changes of the receivers' average remaining energy $E_{sa}$ with CDC and RRC algorithms over the different number of receivers are depicted in Fig. \ref{powernum}.

For the CDC scheduling algorithm, $E_{sa}$ decreases with the increase of $N_r$ in Fig. \ref{powernum}. In addition, from Fig. \ref{Fig3}, given the same transmission distance, the output electric power increases as the input electric power increases. Thus, $E_{sa}$ increases as $P_s$ increases for same $N_r$. For example, when the number of receivers is $10$, $E_{sa}$ of $200$W is greater than that of $150$W, $100$W, and $50$W.

Changes of $E_{sa}$ with different $P_s$ for the RRC scheduling algorithm in Fig. \ref{powernum} is similar to that for the CDC algorithm. Given the same $N_r$ and $P_s$, $E_{sa}$ of the CDC algorithm is higher than that of the RRC algorithm. Since the output electric power $P_e$ is smaller than $1$W when $P_s$ is $50$W from Fig. \ref{Fig3}, $E_{sa}$ of two algorithms with $50$W $P_s$ are close.
\begin{figure}[!t]
    \centering
    \includegraphics[scale=0.6]{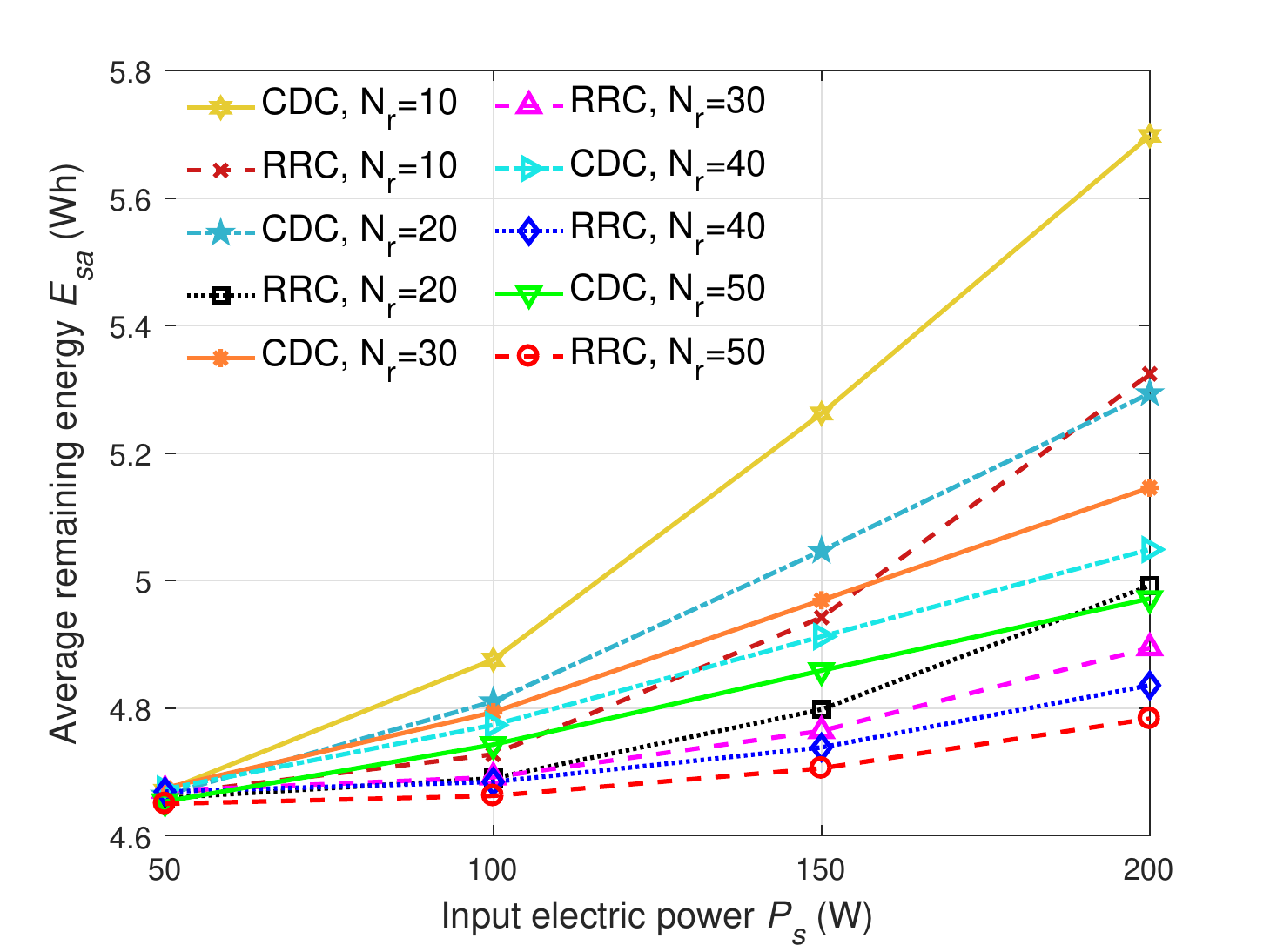}
	\caption{Average remaining energy versus input electric power (different charging duration).}
    \label{numpower}
\end{figure}

Changes of $E_{sa}$ under different input electric power $P_s$ with the different number of receivers $N_r$ are presented in Fig. \ref{numpower}. For the CDC and RRC algorithms, $E_{sa}$ increases gradually with the increase of $P_s$. When $P_s$ is $50$W, $E_{sa}$ is small. Besides, since the denominator of equation \eqref{eq10} increases when increasing the number of receivers, $E_{sa}$ decreases as $N_r$ increases. Compared with the RRC scheduling algorithm, $E_{sa}$ of the CDC scheduling algorithm is always greater with the same $P_s$ and $N_r$.

\begin{figure}[!t]
    \centering
    \includegraphics[scale=0.6]{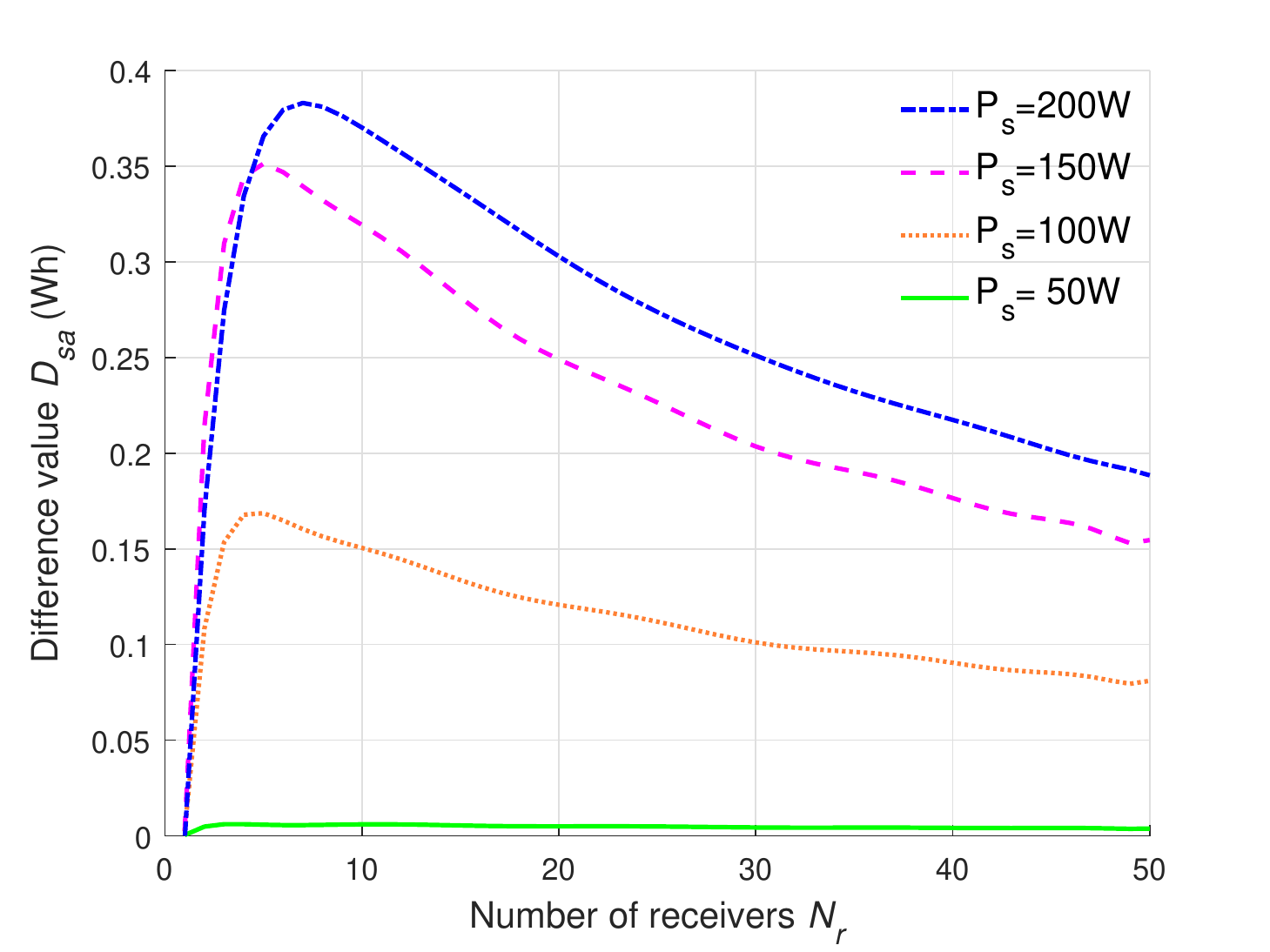}
	\caption{$E_{sa}$ difference value between CDC and RRC versus the number of receivers (different input electric power).}
    \label{differencepower}
\end{figure}

The $E_{sa}$ difference value $D_{sa}$ between the two algorithms under the different number of receivers over diverse $P_s$ is exhibited in Fig. \ref{differencepower}. $D_{sa}$ increases sharply first, then decreases gradually over the number of receivers. In addition, given $N_r$, $D_{sa}$ increases as the input electric power increases. Thus, the performance of the CDC algorithm will get better with higher input electrical power and a reasonable number of receivers.

Therefore, when increasing the input electric power of the RBC transmitter and limiting the number of receivers within the coverage to a reasonable range, the performance of the CDC algorithm can be more superior. In addition, the performance of the CDC algorithm is better than that of the RRC algorithm with the same charging duration, different input electric power and number of receivers.

\subsection{Summary}\label{}

Based on the above performance evaluation and comparison, the analysis results related to the performance of the CDC scheduling algorithm are summarized as follows:

1) The performance of the CDC algorithm improves with longer charging duration, higher input electric power, and reasonable number of receivers within the RBC transmitter's coverage.

2) Compared with the RRC algorithm, the mobile devices can get higher average remaining energy by using CDC algorithm with the same input electric power and charging duration. That is, the effect of energy transmission channel and the receivers' state-of-charge on charging performance of the RBC system is obvious.

3) To improve the performance of the CDC algorithm, the measures that can be taken in the RBC system should include: i) prolonging the charging duration, and, ii) increasing the input electric power, and, iii) limiting the number of receivers within the coverage to a reasonable number.

\section{Conclusions}\label{Section5}
In the Resonant Beam Charging (RBC) system, the transmitter's coverage is different with different input electric power and Field of View (FOV). The wireless energy transmission channels between the transmitter and the receivers may vary. In addition, the receivers' charging power is related to the transmitter's input electric power and the distance to the transmitter. To extend the battery life of the receivers, the remaining energy of all receivers should be maximized. The Channel-Dependent Charge (CDC) scheduling algorithm is proposed to control the receivers' charging power, order, and duration.

Based on the scheduling principle of the CDC algorithm, each receiver is assigned a scheduling coefficient, which is determined by the remaining energy and the energy transmission channel of receiver. The receiver with the minimum scheduling coefficient is selected to be charged first per time slot. The RBC transmission channel model, the transmitter's coverage, and the receiver dynamic moving model are quantified to implement the CDC algorithm. Depending on the performance analysis, the mobile devices can get better service with the CDC algorithm than the Round-Robin Charge (RRC) scheduling algorithm. The methods to improve the CDC performance include: i) prolonging the charging duration, and, ii) increasing the transmitter input electric power, and, iii) limiting the number of receivers charged simultaneously.

Since the wireless transmission channel is complicated in an RBC system, more accurate relationships between the charging power and the transmission channel can be investigated in the future. In addition, due to the RBC transmitter available charging energy is limited, the maximal number of receivers which can be charged simultaneously within an RBC transmitter's coverage can be studied.
%

\bibliographystyle{IEEEtran}
\bibliographystyle{unsrt}
\bibliography{references}

\begin{thebibliography}{10}
\providecommand{\url}[1]{#1}
\csname url@samestyle\endcsname
\providecommand{\newblock}{\relax}
\providecommand{\bibinfo}[2]{#2}
\providecommand{\BIBentrySTDinterwordspacing}{\spaceskip=0pt\relax}
\providecommand{\BIBentryALTinterwordstretchfactor}{4}
\providecommand{\BIBentryALTinterwordspacing}{\spaceskip=\fontdimen2\font plus
\BIBentryALTinterwordstretchfactor\fontdimen3\font minus
  \fontdimen4\font\relax}
\providecommand{\BIBforeignlanguage}[2]{{%
\expandafter\ifx\csname l@#1\endcsname\relax
\typeout{** WARNING: IEEEtran.bst: No hyphenation pattern has been}%
\typeout{** loaded for the language `#1'. Using the pattern for}%
\typeout{** the default language instead.}%
\else
\language=\csname l@#1\endcsname
\fi
#2}}
\providecommand{\BIBdecl}{\relax}
\BIBdecl

\bibitem{sample2011analysis}
A.~P. Sample, D.~T. Meyer, and J.~R. Smith, ``Analysis, experimental results,
  and range adaptation of magnetically coupled resonators for wireless power
  transfer,'' \emph{IEEE Trans. Ind. Electron.}, vol.~58, no.~2, pp. 544--554,
  Feb. 2011.

\bibitem{zhao2015wireless}
N.~Zhao, F.~R. Yu, and V.~C. Leung, ``Wireless energy harvesting in
  interference alignment networks,'' \emph{IEEE Commun. Mag.}, vol.~53, no.~6,
  pp. 72--78, June 2015.

\bibitem{xia2012internet}
F.~Xia, L.~T. Yang, L.~Wang, and A.~Vinel, ``Internet of {T}hings,'' \emph{Int.
  J. Commun. Syst.}, vol.~25, no.~9, pp. 1101--1102, Aug. 2012.

\bibitem{chen2019Learning}
T.~Chen, S.~Barbarossa, X.~Wang, G.~B. Giannakis, and Z.-L. Zhang, ``Learning
  and management for {I}nternet of {T}hings: Accounting for adaptivity and
  scalability,'' \emph{Proc. IEEE}, vol. 107, no.~4, pp. 778--796, Apr. 2019.

\bibitem{wu2014cognitive}
Q.~Wu, G.~Ding, Y.~Xu, S.~Feng, Z.~Du, J.~Wang, and K.~Long, ``Cognitive
  {I}nternet of {T}hings: {A} new paradigm beyond connection,'' \emph{IEEE
  Internet Things J.}, vol.~1, no.~2, pp. 129--143, Apr. 2014.

\bibitem{wang2018power}
H.~Wang, G.~Ding, F.~Gao, J.~Chen, J.~Wang, and L.~Wang, ``Power control in
  {UAV}-supported ultra dense networks: {C}ommunications, caching, and energy
  transfer,'' \emph{IEEE Commun. Mag.}, vol.~56, no.~6, pp. 28--34, June 2018.

\bibitem{guo2016minimizing}
H.~Guo, J.~Liu, Z.~M. Fadlullah, and N.~Kato, ``On minimizing energy
  consumption in {F}i{W}i enhanced {LTE-A} hetnets,'' \emph{IEEE Trans. Emerg.
  Topics Comput.}, vol.~6, no.~4, pp. 579--591, Aug. 2016.

\bibitem{du2017contract}
J.~Du, C.~Jiang, Z.~Han, H.~Zhang, S.~Mumtaz, and Y.~Ren, ``Contract mechanism
  and performance analysis for data transaction in mobile social networks,''
  \emph{IEEE Trans. Netw. Sci. Eng}, vol.~6, no.~2, pp. 103--115, Apr. 2019.

\bibitem{guo2018energy}
H.~Guo, J.~Zhang, J.~Liu, and H.~Zhang, ``Energy-aware computation offloading
  and transmit power allocation in ultra-dense {IoT} networks,'' \emph{IEEE
  Internet Things J.}, vol.~6, no.~3, pp. 4317 -- 4329, June 2018.

\bibitem{cheng20175g}
X.~Cheng, C.~Chen, W.~Zhang, and Y.~Yang, ``5{G}-enabled cooperative
  intelligent vehicular (5{G}en{CIV}) framework: {W}hen {B}enz meets
  {M}arconi,'' \emph{IEEE Intell. Syst.}, vol.~32, no.~3, pp. 53--59, May 2017.

\bibitem{QingqingMET}
Q.~Zhang, G.~Wang, J.~Chen, G.~B. Giannakis, and Q.~Liu, ``Mobile energy
  transfer in {I}nternet of {T}hings,'' \emph{IEEE Internet Things J.}, vol.~6,
  no.~5, pp. 9012--9019, Oct. 2019.

\bibitem{panigrahi2001battery}
D.~Panigrahi, C.~Chiasserini, S.~Dey, R.~Rao, A.~Raghunathan, and K.~Lahiri,
  ``Battery life estimation of mobile embedded systems,'' in \emph{Intl. Conf.
  on VLSI Design}, Bangalore, India, Jan. 3-7, 2001, pp. 57--63.

\bibitem{chi2019energy}
K.~Chi, Z.~Chen, K.~Zheng, Y.-h. Zhu, and J.~Liu, ``Energy provision
  minimization in wireless powered communication networks with network
  throughput demand: {TDMA} or {NOMA}?'' \emph{IEEE Trans. Commun.}, vol.~67,
  no.~9, pp. 6401 -- 6414, Sept. 2019.

\bibitem{sun2017coordinated}
W.~Sun and J.~Liu, ``Coordinated multipoint-based uplink transmission in
  {Internet of Things} powered by energy harvesting,'' \emph{IEEE Internet
  Things J.}, vol.~5, no.~4, pp. 2585--2595, Aug. 2017.

\bibitem{zhao2017joint}
M.-M. Zhao, Q.~Shi, Y.~Cai, and M.-J. Zhao, ``Joint transceiver design for
  full-duplex cloud radio access networks with {SWIPT},'' \emph{IEEE Trans.
  Wireless Commun.}, vol.~16, no.~9, pp. 5644--5658, Sept. 2017.

\bibitem{deng2018multisource}
F.~Deng, X.~Yue, X.~Fan, S.~Guan, Y.~Xu, and J.~Chen, ``Multisource energy
  harvesting system for a wireless sensor network node in the field
  environment,'' \emph{IEEE Internet Things J.}, vol.~6, no.~1, pp. 918--927,
  Feb. 2019.

\bibitem{li2017multiuser}
H.~Li, C.~Huang, and S.~Cui, ``Multiuser gain in energy harvesting wireless
  communications,'' \emph{IEEE Access}, vol.~5, pp. 10\,052--10\,061, June
  2017.

\bibitem{Hui2014}
S.~Y.~R. Hui, W.~Zhong, and C.~K. Lee, ``A critical review of recent progress
  in mid-range wireless power transfer,'' \emph{IEEE Trans. Power Electron.},
  vol.~29, no.~9, pp. 4500--4511, Sept. 2014.

\bibitem{lu2016wireless}
X.~Lu, P.~Wang, D.~Niyato, D.~I. Kim, and Z.~Han, ``Wireless charging
  technologies: Fundamentals, standards, and network applications,'' \emph{IEEE
  Commun. Surveys Tuts.}, vol.~18, no.~2, pp. 1413--1452, 2016.

\bibitem{david20186g}
K.~David and H.~Berndt, ``6{G} vision and requirements: Is there any need for
  beyond 5{G}?'' \emph{IEEE Veh. Technol. Mag.}, vol.~13, no.~3, pp. 72--80,
  Sept. 2018.

\bibitem{wirelesstechniques}
X.~Lu, D.~Niyato, P.~Wang, D.~I. Kim, and Z.~Han, ``Wireless charger networking
  for mobile devices: {F}undamentals, standards, and applications,'' \emph{IEEE
  Wireless Commun.}, vol.~22, no.~2, pp. 126--135, Apr. 2015.

\bibitem{electromagnetic}
A.~Costanzo, M.~Dionigi, D.~Masotti, M.~Mongiardo, G.~Monti, L.~Tarricone, and
  R.~Sorrentino, ``Electromagnetic energy harvesting and wireless power
  transmission: {A} unified approach,'' \emph{Proc. IEEE}, vol. 102, no.~11,
  pp. 1692--1711, Nov. 2014.

\bibitem{cheng2016consumer}
X.~Cheng, R.~Zhang, and L.~Yang, ``Consumer-centered energy system for electric
  vehicles and the smart grid,'' \emph{IEEE Intell. Syst.}, vol.~31, no.~3, pp.
  97--101, May 2016.

\bibitem{liu2016dlc}
Q.~Liu, J.~Wu, P.~Xia, S.~Zhao, W.~Chen, Y.~Yang, and L.~Hanzo, ``Charging
  unplugged: {W}ill distributed laser charging for mobile wireless power
  transfer work?'' \emph{IEEE Veh. Technol. Mag.}, vol.~11, no.~4, pp. 36--45,
  Dec. 2016.

\bibitem{fang2018}
W.~Fang, Q.~Zhang, Q.~Liu, J.~Wu, and P.~Xia, ``Fair scheduling in resonant
  beam charging for {I}o{T} devices,'' \emph{IEEE Internet Things J.}, vol.~6,
  no.~1, pp. 641--653, Feb. 2019.

\bibitem{Qing2017}
Q.~Zhang, X.~Shi, Q.~Liu, J.~Wu, P.~Xia, and Y.~Liao, ``Adaptive distributed
  laser charging for efficient wireless power transfer,'' in \emph{IEEE Veh.
  Technol. Conf.}, Toronto, Canada, Sept. 24-27, 2017, pp. 1--5.

\bibitem{zhang2018distributed2}
Q.~Zhang, W.~Fang, Q.~Liu, J.~Wu, P.~Xia, and L.~Yang, ``Distributed laser
  charging: {A} wireless power transfer approach,'' \emph{IEEE Internet Things
  J.}, vol.~5, no.~5, pp. 3853--3864, Oct. 2018.

\bibitem{xiong2018tdma}
M.~Xiong, M.~Liu, Q.~Zhang, Q.~Liu, J.~Wu, and P.~Xia, ``T{DMA} in adaptive
  resonant beam charging for {I}o{T} devices,'' \emph{IEEE Internet Things J.},
  vol.~6, no.~1, pp. 867--877, Feb. 2019.

\bibitem{fov}
Wi-charge, ``Reference integrations,'' [Onilne]. Available:
  \url{http://www.wi-charge.com/product_category/reference-integrations/}, June
  2018.

\bibitem{zeng2019energy}
Y.~Zeng, J.~Xu, and R.~Zhang, ``Energy minimization for wireless communication
  with rotary-wing {UAV},'' \emph{IEEE Trans. Wireless Commun.}, vol.~18,
  no.~4, pp. 2329--2345, Apr. 2019.

\bibitem{wang2018channel}
W.~Wang, Q.~Zhang, H.~Li, M.~Liu, X.~Liao, and Q.~Liu, ``Wireless energy
  transmission channel modeling in resonant beam charging for {I}o{T}
  devices,'' \emph{IEEE Internet Things J.}, vol.~6, no.~2, pp. 3976--3986,
  Apr. 2019.

\bibitem{yu2013general}
S.~Yu, R.~Doss, W.~Zhou, and S.~Guo, ``A general cloud firewall framework with
  dynamic resource allocation,'' in \emph{IEEE Intl. Conf. on Commun.},
  Budapest, Hungary, June 9-13, 2013, pp. 1941--1945.

\bibitem{cheng2015d2d}
X.~Cheng, L.~Yang, and X.~Shen, ``{D}2{D} for intelligent transportation
  systems: {A} feasibility study,'' \emph{IEEE Trans. Intell. Transp. Syst.},
  vol.~16, no.~4, pp. 1784--1793, Aug. 2015.

\bibitem{fang2018earning}
W.~Fang, Q.~Zhang, M.~Liu, Q.~Liu, and P.~Xia, ``Earning maximization with
  quality of charging service guarantee for {I}o{T} devices,'' \emph{IEEE
  Internet Things J.}, vol.~6, no.~1, pp. 867--877, Feb. 2019.

\bibitem{aziz2014simulation}
M.~S. Aziz, S.~Ahmad, I.~Husnain, A.~Hassan, and U.~Saleem, ``Simulation and
  experimental investigation of the characteristics of a {PV}-harvester under
  different conditions,'' in \emph{Intl. Conf. on Energy Syst. Polic.},
  Islamabad, Pakistan, Nov. 21-23, 2014, pp. 1--8.

\bibitem{xiong2019resonant}
M.~Xiong, Q.~Liu, G.~Wang, G.~B. Giannakis, and C.~Huang, ``Resonant beam
  communications: Principles and designs,'' \emph{IEEE Commun. Mag.}, vol.~57,
  no.~10, pp. 34--39, Oct. 2019.

\bibitem{koechner2013solid}
Koechner and Walter, \emph{Solid-state {L}aser {E}ngineering}.\hskip 1em plus
  0.5em minus 0.4em\relax Springer-Verlag, 1976.

\bibitem{penzkofer1988solid}
A.~Penzkofer, ``Solid state lasers,'' \emph{Pro. Quan. Electron.}, vol.~12,
  no.~4, pp. 291--427, 1988.

\end{thebibliography}

\end{document}